\documentclass[aps,pre,reprint]{revtex4-1}
\usepackage{times}		
\usepackage{booktabs}
\usepackage{placeins}

\usepackage{amssymb,amstext,amsmath, epsf}
\usepackage{graphicx,graphics,epsfig,wrapfig}
\usepackage{listings}
\usepackage{float}
\usepackage{esint}

\usepackage{color}
\usepackage[english]{babel}
\usepackage[T1]{fontenc}
\usepackage[utf8]{inputenc}

\def\apj{{ApJ}}
\def\apjl{{ApJL}}

\def\mnras{{MNRAS}}
\def\nat{{Nature Physics}}
\def\prl{{PRL}}

\def\pre{{PRE}}

\def\aap{{Astronomy and Astrophysics}}

\begin{document}

\title{On the Quasicollisionality of Plasmas with Small-Scale Electric Turbulence}

\author{Brett D. Keenan}
\email{bdkeenan@ku.edu}
\affiliation{Department of Physics and Astronomy, University of Kansas, Lawrence, KS 66045}
\author{Mikhail V. Medvedev}
\affiliation{Department of Physics and Astronomy, University of Kansas, Lawrence, KS 66045}


\begin{abstract}
Chaotic electromagnetic fields are common in many relativistic plasma environments, where they can be excited by instabilities on kinetic spatial scales. When strong electric fluctuations exist on sub-electron scales, they may lead to small-angle, stochastic deflections of the electrons' pitch-angles. Under certain conditions, this closely resembles the effect of Coulomb collisions in collisional plasmas. The electric pitch-angle diffusion coefficient acts as an effective collision -- or ``quasi-collision'' -- frequency. We show that quasi-collisions may radically alter the expected radiative transport properties of candidate plasmas. In particular, we consider the quasi-collisional generalization of the classical Faraday effect.
\end{abstract}

\maketitle

\section{Introduction}
\label{s:intro}

Electromagnetic turbulence has enormous significance to many astrophysical plasmas. It is believed to play a significant role, for example, in the mediation of supernova shocks and $\gamma$-ray bursts \citep{medvedev09b, medvedev06, reynolds12, kamble+14}. In other high-energy-density (HED) environments, such as the laboratory setting, the manipulation and control of electromagnetic turbulence is crucial to fusion energy science and inertial confinement fusion (ICF) \citep{ren04, tatarakis03}. Stochastic electromagnetic fields are important to laboratory astrophysics laser-plasma experiments as well \citep{huntington15,park15}. In many cases, these fields reside at extremely small spatial scales. For this reason, the consideration of particle transport and radiation production in these plasmas requires a special treatment.

In our previous work, Ref. \citep{keenan15c}, we showed that sub-Larmor-scale (``small-scale'') magnetic turbulence induces particle dynamics reminiscent of binary Coulomb interactions. A stochastic magnetic field is deemed ``sub-Larmor-scale'', with respect to a sub-population of electrons, if the electrons' {\em effective} Larmor radius, $r_L \equiv \gamma_e\beta m_e c^2/e \langle\delta B^2\rangle^{1/2}$ is greater than, or comparable to, the magnetic correlation length, $\lambda_B$, i.e., $\lambda_B\lesssim r_L$. Here $\beta=v/c$ is the dimensionless particle velocity, $\langle\delta B^2\rangle^{1/2}$ is the rms value of the fluctuating magnetic field, $m_e$ is the electron mass, $c$ is the speed of light, $e$ is the electric charge, and $\gamma_e$ is the electron's Lorentz factor. 

The random small-angle deflections of electrons, caused by small-scale magnetic fields, leads to an effective collisionality. This is because the magnetic deflections are always transverse to the direction of motion; thus, they resemble, in the small deflection angle regime, random Coulomb deflections. The magnetic ``quasi-collision'' frequency is equal to the (small-angle) pitch-angle diffusion coefficient. In this work, we will show that this concept may be readily generalized to the case of relativistic electrons moving through electric turbulence. Although electric fields, in general, lead to both transverse and parallel accelerations, if the electron is moving sufficiently fast, then the transverse change in momentum will be far greater than the parallel impulse -- hence, the electron's motion will approximately match the purely transverse deflections seen in the magnetic case.

In this work, we will properly define the ``small-scale'' regime for pure electric turbulence, and we will derive the pitch-angle diffusion coefficient -- i.e.\ the quasi-collision frequency. We will, furthermore, investigate the consequences this realization of quasi-collisionality has for the dielectric properties of plasmas with these small-scale electric fields. In particular, we will explore Faraday rotation in magnetized plasmas with small-scale electric fluctuations. We will show that strong small-scale fields, and the accompanying strong quasi-collisionality, may radically modify the Faraday rotation effect in these plasmas -- possibly leading, in fact, to negative rotation measures.

Finally, we will consider the environments best suited for the physical realization of these effects. We will argue that plasmas with mildly relativistic electrons, but yet non-relativistic ions, are likely required. We will show that plasmas with short-wavelength ion acoustic turbulence present a very likely candidate.

The rest of the paper is organized as follows. Section \ref{s:analytic} briefly reviews the analytic theory of pitch-angle diffusion in small-scale random electric fields. We then show that the pitch-angle diffusion coefficient, itself, acts as an effective collision frequency. In Section \ref{s:coll_eqs}, we explore the implications for electromagnetic wave propagation in magnetized plasmas with high quasi-collisionality. Next, Section \ref{s:physical} considers environments favorable for the realization of these effects. Finally, Section \ref{s:concl} is the conclusions. We use cgs units throughout the paper.

\section{Analytic Theory}
\label{s:analytic}

Suppose an electron test particle is moving, with speed, $v$, through an external random electric field. We will assume that the electric field fluctuates very slowly -- such that the particle dynamics, on relevant time-scales, are largely unaffected by the field's time-variability.

For ``small-scale'' turbulence, the principal time-scale which governs particle transport is the time to transit a single electric field correlation length, $\lambda_E^t$ -- where the ``$t$'' superscript indicates that the correlation length is specified along the path with a ``transverse'' component of the electric field. If the transit time, $\tau_E^t \sim \lambda_E^t/v$, is much less than the field-variability time-scale, $\Omega_r^{-1}$, then we may treat the electric field as approximately time-independent. 

To proceed, we it be helpful to discuss the radiation produced by an electron moving through an external random field. In a random electromagnetic field, the acceleration occurs principally along the extent of a correlation length. Since we assume that the electron is moving ultrarelativistically, it will undergo a slight deflection, $\delta\alpha_E$, as it traverses $\lambda_E^t$. If $\delta\alpha_E$ is much less than the radiation beaming angle, $\Delta{\theta} \sim 1/\gamma_e$, then the electron will move approximately rectilinearly, undergoing only slight random deflections along its path. In this case, the radiation will be beamed along the extent of the electron's relatively fixed direction of motion. An observer on axis would, therefore, see a signal for the electron's entire trajectory. Furthermore, the radiation spectrum will be wholly determined by the statistical properties of the underlying acceleration mechanism \citep{landau75} -- which, in this case, is an electric field. The electron emits radiation in the small-angle jitter regime when the acceleration mechanism is a random (static) magnetic field \citep{medvedev00, medvedev06, medvedev11, RK10, TT11, keenan13, keenan15}. Similarly, the radiation produced by ultrarelativistic electrons moving through electrostatic turbulence, in the small-scale regime, is nearly identical; for this reason, it is considered a subclass of small-angle jitter radiation \citep{teraki14, keenan16}.

We have previously shown that these random deflections initiate pitch-angle diffusion in sub-Larmor-scale magnetic turbulence, and that this diffusion coefficient is intimately related to the radiation spectrum \citep{keenan13, keenan15}. Later, we showed that this relation holds for small-scale electric turbulence, as well \citep{keenan16}. 

We consider an electric field as ``small-scale'' if: 
\begin{subequations}     
\begin{align}
\Omega_r^{-1} \gg \tau_E^t,  \\  
\Delta{\theta} \gg \delta\alpha_E.
\end{align}
\label{small_scale_def}
\end{subequations}
Since the electron is moving ultrarelativistically, the component of its acceleration transverse to its direction of motion will be far larger than the parallel component. Thus, its motion occupies the small deflection angle regime when $\Delta{\theta} \gg \delta\alpha_E$ -- which is the reason its radiation spectrum resembles the jitter spectrum. Additionally, transverse accelerations leave the particle's kinetic energy fixed. For this reason, we will assume a constant $v$. 

Next, since the deflections are assumed to be small, $\delta\alpha_E \sim \Delta{p}_t/p$ -- where $p = \gamma_em_ev$ is the kinetic momentum of the electron, and $\Delta{p}_t$ is the change in its transverse momentum. Since $\Delta{p}_t/\tau_E \sim eE_t$, where $E_t$ is the component of the electric field perpendicular (transverse) to the electron's direction of motion, $\Delta{p}_t/p \sim eE_t/\gamma_e{m_e}v$; thus:
\begin{equation}
\delta\alpha_E \sim \frac{eE_t}{\gamma_e{m_e}v}\tau_E.
\label{alph_def}
\end{equation}
Consequently, since $D^\text{elec.}_{\alpha\alpha} \sim \delta\alpha_E^2/\tau_E$, the electric diffusion coefficient must be:
\begin{equation}
D^\text{elec.}_{\alpha\alpha} = \frac{\lambda_E^t}{\gamma_e^2c\beta^3}\langle {\Omega_E^\perp}^2 \rangle,
\label{Daa}
\end{equation}
where:
\begin{equation}
\Omega_E^\perp \equiv eE_t/m_ec.
\label{oe_def}
\end{equation}

\subsection{Pitch-angle Diffusion as Effective Collisionality}
\label{s:effective}

The small-angle electric deflections are analogous to electron-ion collisional deflections in a number of ways; they both approximately conserve particle energy, and they both induce deflections that are approximately transverse to the electron's initial velocity.

Where the two effects differ, however, is in the nature of the stochasticity. In an idealized scenario, an electron in a collisional plasma is continuously deflected by ions along its trajectory. In contrast, an electron moving through small-scale electric turbulence is deflected on a characteristic spatial scale of finite length: the correlation length. Thus, the two descriptions are only equivalent on a coarse-graining. Indeed, the electron motion in small-scale turbulence resembles electron-ion collisions only on spatial scales greater than -- or similar to -- the electric correlation length. 

Thus, we must require that:
\begin{equation}
L \gtrsim \lambda_E^t,
\label{size_def}
\end{equation}
where $L$ is the characteristic length scale of the system. With regard to the propagation properties of plasmas, this dimension is on the order of the wave packet size. For pure plane waves, however, $L$ is unlimited. 

Next, we may infer this effective collision frequency directly from Eq.\ (\ref{Daa}). The pitch-angle deflections are assumed to be small. Thus, at $\tau_c$, the following condition must hold:
\begin{equation}
D^\text{elec.}_{\alpha\alpha}\tau_c \sim 1.
\label{coll_def_mag}
\end{equation}
Therefore, $D^\text{elec.}_{\alpha\alpha}$ must be the effective ``collision'' frequency.

\subsection{A Phenomenological Definition of the Quasi-Collision Frequency}
\label{s:phenomenal}

Estimating $D^\text{elec.}_{\alpha\alpha}$, in real plasmas, may be difficult since it depends upon the small-scale correlation length -- a quantity which requires knowledge of the electric spectral distribution to obtain. In principle, if the nature of the instability which produces the electric fluctuations is known, then we may produce a rough estimate of the characteristic spatial scales which ultimately set the correlation length. However, in many cases, the type of turbulent fluctuations may not be known; hence, an {\it a priori} estimate of the electric spectrum may not be available.

In Ref. \citep{keenan15c}, we showed that the emission coefficient, $j_\omega$ --- which is the radiant power per unit frequency per unit volume per unit solid-angle --- is directly proportional to the (magnetic) quasi-collision frequency; this is akin to the Bremsstrahlung radiation equivalent which defines an effective electron-ion collision frequency \citep{bekefi66}:
\begin{equation}
j^\textrm{Brems}_\omega = \text{Re}[n]\left(\frac{\omega_\text{pe}^2k_BT_e}{8\pi^3c^3}\right)\nu_{ei},
\label{comb_emiss}
\end{equation}
where $\text{Re}[n]$ is the real part of the plasma's index of refraction, $\omega_\text{pe}= \sqrt{4\pi n_e{e^2}/m_e}$ is the (non-relativistic) electron plasma frequency, $T_e$ is the electron temperature, $n_e$ is the electron number density, $k_B$ is the Boltzmann constant, and $\nu_{ei}$ is an electron-ion collision frequency. The jitter equivalent follows a similar pattern; to wit \citep{keenan15c}:
\begin{equation}
j^\textrm{jitter}_\omega \sim \left(\frac{m_e\omega_\text{pe}^2}{24\pi^2c}\right)\gamma_e^2e^2\beta^2D^\text{mag.}_{\alpha\alpha},
\label{jitt_emiss}
\end{equation} 
where:
\begin{equation}
D^\text{mag.}_{\alpha\alpha} \propto \frac{\lambda_B^t}{\gamma_e^2c\beta}\langle \Omega_B^2 \rangle,
\label{Daa_mag}
\end{equation}
is the magnetic pitch-angle diffusion coefficient. Comparing Eqs.\ (\ref{Daa_mag}) and (\ref{Daa}), it is clear that the main difference is the $\beta$-factor in the denominator.

Since relativistic electrons in small-scale electric turbulence, also, emit jitter radiation \citep{teraki14, keenan16}, we may do the same here. First, we must consider the total radiated power of the electron. To this end, we use the general Larmor formula, which is given by \citep{jackson99}:
\begin{equation}
P_\text{tot.} = \frac{2e^2\gamma_e^6}{3c}\left[\dot{\boldsymbol \beta}^2 - ({\boldsymbol \beta} \times \dot{\boldsymbol \beta})^2\right].
\label{Larmor_power}
\end{equation} 
For purely electric fields, we have the acceleration:
\begin{equation}
\dot{\boldsymbol \beta} = -\frac{1}{\gamma_e}\left[{\boldsymbol\Omega}_E - {\boldsymbol\beta}\left({\boldsymbol\beta}\cdot{\boldsymbol\Omega}_E\right) \right].
\label{dvdt}
\end{equation}
However, we are assuming that the transverse acceleration dominates, hence:
\begin{equation}
\dot{\boldsymbol \beta} \approx -\frac{1}{\gamma_e}{\boldsymbol\Omega}_E^\perp,
\label{dvdt2}
\end{equation}
and, therefore:
\begin{equation}
P_\textrm{tot}^\textrm{jitter}  \approx \frac{2}{3}{c}r_e^2\gamma_e^2E_t^2,
\label{jitter_power}
\end{equation} 
where $r_e = e^2/m_e c^2$ is the classical electron radius. Next, the small-angle jitter radiation spectrum has a characteristic frequency known as the jitter frequency,
\begin{equation}
\omega_{j} = \gamma_e^2k_E\beta{c},
\label{jitter_freq}
\end{equation} 
where $k_E$ is the dominant wave number of the (small-scale) turbulent fluctuations. Next, we may write the spectral power for a single electron as:
\begin{equation}
P_\textrm{jitter}(\omega) \equiv \frac{dP}{d\omega} \sim \frac{P_\textrm{tot}^\textrm{jitter}}{\omega_j}.
\label{spec_pow_def}
\end{equation} 
Substitution of Eq.\ (\ref{jitter_freq}) into Eq.\ (\ref{jitter_power}), results in the expression:
\begin{equation}
P_\textrm{jitter}(\omega) \sim \frac{2}{3}\lambda_E^t \beta^{-1} \left(\frac{e^4}{m_e^2c^4}\right)E_t^2,
\label{spec_pow}
\end{equation} 
where the relation, $k_E^{-1} \sim \lambda_E^t$, has been employed \citep{keenan15}. Comparing this result to Eq.\ (\ref{Daa}), we find that the power spectrum is directly proportional to the pitch-angle diffusion coefficient:
\begin{equation}
P_\textrm{jitter}(\omega) \sim \frac{2}{3}\frac{e^2}{c}\gamma_e^2\beta^2D^\text{elec.}_{\alpha\alpha},
\label{spec_pow_diff}
\end{equation} 
which is equivalent to the magnetic expression \citep{keenan15c}.

Finally, if we assume isotropic emission by all plasma electrons, then the jitter emission coefficient may be obtained from Eq.\ (\ref{spec_pow_diff}) with the multiplication of $n_e/4\pi$. Thus:
\begin{equation}
j^\textrm{jitter}_\omega = \left(\frac{m_e\omega_\text{pe}^2}{24\pi^2c}\right)\gamma_e^2e^2\beta^2D^\text{elec.}_{\alpha\alpha}.
\label{jitt_emiss}
\end{equation} 
Thus, Eq.\ (\ref{jitt_emiss}) -- once again -- provides an attractive phenomenological definition for the ``jitter'' collision frequency, which may be obtained directly from the small-angle jitter radiation emission coefficient.

\section{``Quasi-collisional'' Faraday Effect}
\label{s:coll_eqs}

Faraday rotation is the result of magnetically-induced birefringence in a dielectric medium. When light propagates through a plasma -- parallel to a uniform magnetic field -- the left- and right-circular polarizations have different indices of refraction. Consequently, a linearly polarized wave would suffer a rotation of its polarization axis as it traverses such a medium, since any linear polarization may be envisioned as the superposition left- and right-circular polarizations.

In Ref.\ \citep{keenan15c}, we showed that magnetic-induced quasi-collisionality alters the expected form of this Faraday rotation, $\Delta\Psi$, for magnetized plasmas. The ``collisionless'' Faraday rotation is given by the, well-known, expression:
\begin{equation}
\Delta\Psi_\text{collisionless} = \lambda^2 R_M,
\label{fara_int}
\end{equation} 
where $\lambda$ is the electromagnetic radiation wavelength, and the standard {\it rotation measure} is:
\begin{equation}
R_M \equiv \frac{e^3}{2\pi m_e^2c^2} \int n_e(z)B_\parallel(z) \textrm{d}z,
\label{R_M_def}
\end{equation} 
where $B_{\parallel}(z)$ is the component of the magnetic field, at $z$, parallel to electromagnetic wave-vector. The ``collisionless'' and ``quasi-collisional'' expressions have the ratio \citep{keenan15c}:
\begin{equation}
\frac{\Delta\Psi_\text{quasi-collisional}}{\Delta\Psi_\text{collisionless}} \simeq \frac{\left(1 - Z^2\right)}{\left(1 + Z^2\right)^2},
\label{discrep_rat}
\end{equation} 
where $Z \equiv \nu_\text{eff}/\omega$ is the normalized quasi-collision frequency, $\nu_\text{eff}$ is the (effective) quasi-collision frequency, and $\omega = 2\pi{c}/\lambda$. Additionally, quasi-collisions lead to absorption, with an absorption coefficient, $\alpha_\textrm{absp}^\textrm{Farad}$, given by \citep{keenan15c}:
\begin{equation}
\alpha_\textrm{absp}^\textrm{Farad} \equiv -\frac{\omega_\text{pe}^2\nu_\text{eff}}{c\left(\omega^2 + {\nu_\text{eff}}^2\right)}\left[1 \mp \frac{2\Omega_\text{ce}\omega}{\left(\omega^2 + {\nu_\text{eff}}^2\right)}\right],
\label{absp_def2}
\end{equation} 
where $\Omega_\text{ce} \equiv eB_\parallel/m_ec$ and the $\pm$ sign refers to the right- and left-circular polarizations, respectively.

Eq.\ (\ref{fara_int}) formally holds for a non-relativistic, ``cold'' plasma with $\omega \gg \Omega_\text{ce}$ and $\omega^3 \gg \omega_\text{pe}^3$. Consequently, it fails for the scenario considered here, since Eq.\ (\ref{small_scale_def}) implies that the electron population is relativistic.

Fortunately, the Faraday expression for relativistic velocities is a straightforward generalization of Eq.\ (\ref{fara_int}) \citep{shcherbakov08}:
\begin{equation}
\Delta\Psi_\text{collisionless} \approx \frac{K_0(\sigma)}{K_2(\sigma)}\lambda^2 R_M,
\label{fara_int_rel}
\end{equation} 
where $\sigma \equiv m_ec^2/k_BT_e$ and $K_j(x)$ is a modified Bessel function of the second-kind. Thus, under the first-order substitution rule for including the effects of collisions, i.e.\ $\omega \rightarrow \omega + i\nu_\text{eff}$, Eq.\ (\ref{discrep_rat}) will hold for the relativistic regime as well. Likewise,
\begin{equation}
\alpha_\textrm{absp}^\textrm{Farad} \rightarrow \frac{K_0(\sigma)}{K_2(\sigma)}\alpha_\textrm{absp}^\textrm{Farad}.
\label{absorp_rel}
\end{equation} 
Finally, we will assume that the small-scale electric fluctuations are predominantly along the direction of the ambient magnetic field, ${\bf B}$. This assumption allows us to disregard additional complications, such as diffusion induced by ``E cross B'' drifts.

\begin{figure}
\includegraphics[angle = 0, width = 1\columnwidth]{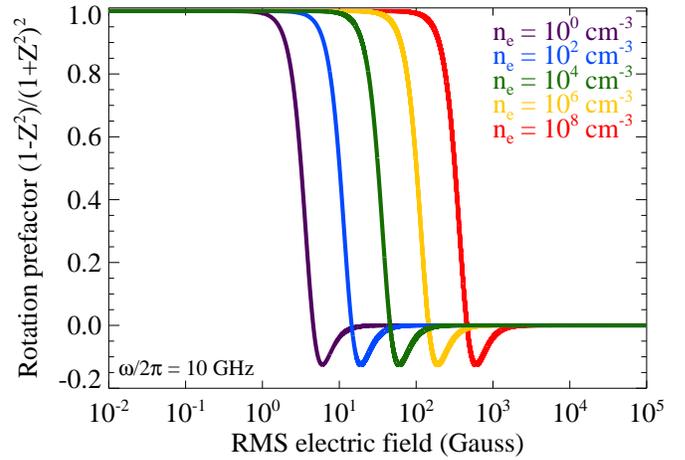}
\caption{(Color online) Normalized Faraday rotation angle -- i.e.\ Eq.\ (\ref{discrep_rat}) -- vs. the electric fluctuation strength for various electron densities. The electron number densities are (from right to left): $n_e = 1$, $10^{2}$, $10^{4}$, $10^{6}$, and $10^{8} \ \text{cm}^{-3}$; the electron temperature is $0.511 \ \text{MeV}$ (or $\gamma_e \sim 2$), and $\omega/2\pi = 10 \ \text{GHz}$. These curves exhibit a universal feature; namely, the rotation angle will reverse sign when the electric fluctuation field is sufficiently strong. As $\langle E^2 \rangle^{1/2} \rightarrow \infty$, the rotation is completely nullified. }
\label{farad}
\end{figure}

The properties of the curve represented by Eq.\ (\ref{discrep_rat}) are explored in Ref.\ \citep{keenan15c}. In Figure \ref{farad}, we have plotted Eq.\ (\ref{discrep_rat}) as a function of the electric fluctuation strength (the ``rms'' value of the electric field) for mildly relativistic electrons ($\gamma_e \sim 2$). As shown in Ref.\ \citep{keenan16}, the small deflection angle regime holds well even at these mildly relativistic speeds. Five curves appear in Figure \ref{farad}, each differing by the electric correlation length, which is chosen to be equal to the relativistic electron skin-depth, $d_e = c\sqrt{\gamma_e}/\omega_\text{pe}$ (the reasons for this choice will become more apparent in the following section). The electron number densities are: $n_e = 1$, $10^{2}$, $10^{4}$, $10^{6}$, and $10^{8} \ \text{cm}^{-3}$; the electromagnetic wave frequency, $\omega/2\pi$, is $10 \ \text{GHz}$. 

The curves in Figure \ref{farad} exhibit a universal feature: the rotation angle reverses sign when the electric fluctuation field is sufficiently strong. As $\langle E^2 \rangle^{1/2} \rightarrow \infty$, the rotation is completely nullified. Notice that, for typical interstellar densities ($n_e \sim 1 \ cm^{-3}$), a noticeable effect can be seen for electric field strengths $> 1.0 \ \text{G}$.

\begin{figure}
\includegraphics[angle = 0, width = 1\columnwidth]{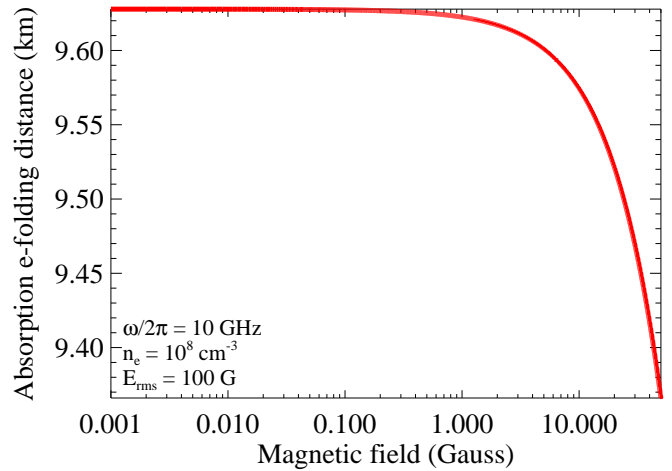}
\caption{(Color online) Quasi-collisional absorption e-folding distance vs. the ambient magnetic field. The instance depicted here is the $n_e = 10^8 \ \text{cm}^{-3}$ case from Figure \ref{farad}. We see that the ``collisional'' absorption occurs on a many kilometer length scale. With $B_0 = 10 \ \text{G}$, the signal would be reduced to a factor of $0.01$ around $44 \ \text{km}$.}
\label{absorp}
\end{figure}

Next, since strong quasi-collisions imply strong ``collisional'' absorption, we must consider the result of Eq.\ (\ref{absorp_rel}) -- which will, in turn, constrain the strength of the ambient magnetic field. In Figure \ref{absorp}, the e-folding distance (i.e.\ $1/\alpha_\textrm{absp}^\textrm{Farad}$) is plotted as function of $B_0$ for the $n_e = 10^8 \ \text{cm}^{-3}$ case from Figure \ref{farad}. We see that the ``collisional'' absorption occurs on a many kilometer length scale. With $B_0 = 10 \ \text{G}$, the signal would be reduced to a factor of $0.01$ around $44 \ \text{km}$. Thus, the limiting factor in the possible observation of the quasi-collisional Faraday effect is this absorption; which is, necessarily, strong when the quasi-collision frequency is large.
In the penultimate section, we will argue that small-scale ion acoustic turbulence may present the ideal realization of electric quasi-collisionality in actual space, astrophysical, and laboratory plasmas. 

\section{Small-Scale Electric Turbulence in Real Plasmas}
\label{s:physical}

For the realization of small-scale electric turbulence, the second condition from Eq.\ (\ref{small_scale_def}) is most difficult to satisfy. Using $\tau_E \sim c/\lambda_E^t$ and $\lambda_E^t \sim k_E^{-1}$, this condition is equivalent to:
\begin{equation}
\frac{k_E^2c^2}{\Omega_r^2} \gg 1.
\label{cond_ult}
\end{equation} 
In accord with Refs.\ \citep{teraki14, keenan16}, we may be inclined to choose ``cold'' electron langmuir waves to mediate the turbulence. The dispersion relation, in this case, would be $\Omega_r = \omega_\text{pe}^\text{rel.}$ -- where $\omega_\text{pe}^\text{rel.} \equiv \omega_\text{pe}/\sqrt{\gamma_e}$, is the relativistic plasma frequency. Thus, Eq.\ (\ref{cond_ult}) would require that:
\begin{equation}
k_Ec \gg \omega_\text{pe}^\text{rel.},
\label{cond_electron}
\end{equation} 
or, equivalently, that the correlation length is smaller than the electron skin-depth. However, this is problematic. The thermally corrected, ultra-relativistic, dispersion relation for electron langmuir waves is\citep{bergman01}:
\begin{equation}
\Omega_r^2 = \frac{\sigma}{3}\omega_\text{pe}^2 + \frac{9}{5}k^2c^2.
\label{langmuir}
\end{equation} 
Thus, the condition that $\Omega_r \simeq \omega_\text{pe}^\text{rel.}$ contradicts Eq.\ (\ref{cond_electron}), since these ``cold'' plasma waves require that $(27/5)k^2c^2 \ll \sigma\omega_\text{pe}^2 \sim \omega_\text{pe}^2/\gamma_e$. Therefore, electron Langmuir turbulence may not self-consistently satisfy all the conditions that we require.

Alternatively, we may consider turbulence mediated by the ion population. In this case, we must be careful to specify electric fluctuations that exist on spatial scales comparable to the Debye length, $\lambda_D$, since the ion time-scales will be sufficiently long enough that electrons will effectively screen out these fields on electron scales; i.e.\ these ``large-scale'' electric fields have very little effect on the electron population.

When electric fluctuations exist on scales smaller than the Debye shielding length, then ``quasi-neutrality'' can be broken. Thus, with $m_i \gg \gamma_em_e$ (where $m_i$ is the ion mass), we require that:
\begin{equation}
k_E\lambda_D \gtrsim 1,
\label{disp_cond}
\end{equation} 
where $\lambda_D = v_\text{the}/\omega_\text{pe}$, and $v_\text{the} \sim c$ is the electron thermal velocity. 

One possible realization of this condition is provided by the very short wavelength branch of the ion acoustic mode. These modes can exist in magnetized plasmas, and we assume for our purposes that the wave-vector is nearly aligned with the direction of the ambient magnetic field. For a non-relativistic, ``cold'' plasma, If $k_E\lambda_D \gtrsim 1$, the ion acoustic mode has the frequency \citep{luo97}:
\begin{equation}
\Omega_r = \frac{\omega_\text{pi}}{1 + (k\lambda_D)^{-2}},
\label{ion_acoustic}
\end{equation} 
where $\omega_\text{pi}$ is the ion plasma frequency. From this, we see that $\Omega_r \approx \omega_\text{pi}$ when $k\lambda_D \gg 1$. Ion acoustic turbulence may be strongly excited when the electron temperature far exceeds the ion temperature \citep{luo97} -- a scenario which is required here, since the ions are being treated as non-relativistic, while the electrons are -- at least -- mildly relativistic (i.e.\ $m_i \gg \gamma_em_e$).

\section{Conclusions}
\label{s:concl}

In our previous work, Ref. \citep{keenan15c}, we investigated ``quasi-collisionality'' induced by small-scale magnetic turbulence in, otherwise, collisionless plasma environments. We found that the pitch-angle diffusion coefficient \citep{keenan13, keenan15} acts as an effective collision frequency. In this work, we extended this concept to small-scale electrostatic turbulence -- in the relativistic regime. 

We derived the quasi-collision frequency -- Eq.\ (\ref{Daa}) -- for relativistic electrons moving through small-scale electric turbulence. Just as the electron-ion (Coulomb) collision frequency may be conveniently defined by the Bremsstrahlung emission coefficient, Eq.\ (\ref{jitt_emiss}) offers a simple phenomenological representation for the quasi-collision frequency in terms of the (electric) jitter emission coefficient.  Jitter radiation may be directly observable in several of these plasma environments. In fact, there is evidence that the (magnetic) jitter radiation from mildly relativistic electrons may be observable in high-intensity solid-density laser plasma experiments \citep{keenan15b}. 

Next, we explored the consequences of high electrically-induced quasi-collisionality for Faraday rotation in magnetized plasmas. We found -- as we did, previously, for magnetically-induced quasi-collisionality -- that the Faraday rotation measure, $R_M$, may obtain negative values, in this case. In fact, as the quasi-collision frequency becomes sufficiently large, $R_M \rightarrow 0$. 

We, furthermore, speculated upon the most likely set of plasma parameters that would allow for direct observation of this, modified, Faraday effect. We found that quasi-collisional absorption may severely limit possible space and astrophysical applications of our model -- since strong quasi-collisionality, also, implies strong ``collisional'' absorption.

Finally, we argued that plasmas with high-frequency, small-scale, ion-acoustic turbulence -- where $T_e \gg T_i$ -- offer the most likely environment in which these effects may be physically realized. Supernova remnant (SNR) shocks, for example, may host ion-acoustic instabilities that may drive the required strong, turbulent fluctuations \citep{dieckmann00}. Acoustic modes have, additionally, been implicated in the phenomenon of pulsar eclipsing, and astrophysical accretion flows where $T_i \neq T_e$ \citep{luo97}. Thus, a number of astrophysical environments may be favorable candidates.

However, owing to the high ``collisional'' absorption that accompanies high quasi-collisionality, the Faraday signature of these plasmas may be completely obscured. For this reason, space and laboratory plasmas may be better suited for the direct observation of this unique signature. Laser-plasmas, specifically, are an attractive candidate -- since sufficiently intense laser pulses can quickly heat an electron population and separate it from an ion background. Such a plasma configuration is especially susceptible to ion-acoustic instabilities.

To conclude, the obtained results suggest that small-scale electric fluctuations conceal a “collisional” signature, which may provide a useful radiative diagnostic tool -- via its strong impact upon Faraday rotation in these plasmas -- for the evaluation of small-scale electric turbulence in laboratory, astrophysical, space and solar plasmas.

\end{document}